\documentstyle[prb,aps]{revtex}

\begin{document}
\title{Observation of an incoherent thermally activated proton hopping process in
calix-[4]-arene by means of anelastic spectroscopy.}
\author{A. Paolone, R. Cantelli}
\address{Istituto Nazionale per la Fisica della Materia and Dipartimento di Fisica,\\
Universit\`{a} di Roma ``La Sapienza``,\\
P.le A. Moro 2, I-00185 Roma, Italy}
\author{R. Caciuffo}
\address{Istituto Nazionale per la Fisica della Materia and Dipartimento di Fisica ed%
\\
Ingegneria dei Materiali e del Territorio, Universit\`{a} di\ Ancona, Via\\
Brecce Bianche, I-60131 Ancona, and INFM, Italy}
\author{A. Arduini}
\address{Dipartimento di Chimica Organica e Industriale, Universit\`{a} di Parma,\\
Parco Area delle Scienze 17/a, I-43100 Parma, Italy}
\maketitle

\begin{abstract}
The anelastic spectrum of calix[4]arene was measured at two different
vibrational frequencies. Three thermally activated peaks were detected. The
lowest temperature peak can be described considering a continous
distribution function of activation energies for the relaxation. This
anelastic peak can be ascribed to a thermally activated hopping process of H
atoms of the OH groups, corresponding to a flip-flop of the OH bond. From
the results of the present study, it seems that anelastic spectroscopy is a
good experimental technique to study atomic motion inside molecules at a
mesoscopic (few molecules) level.
\end{abstract}


\twocolumn

\section{INTRODUCTION}

Calix[{\it n}]arenes are cup-like [l$_{n}$] metacyclophanes which derive
from the condensation of phenols and formaldehyde in different conditions.%
\cite{Gutsche} The bracketed number indicate the number of phenol units (the
atoms inside the rectangle in Fig.1) and hence defines the size of the
macrocycle. The tetramer is named calix[4]arene (see Fig.1) The number of
the phenolic units can vary from 3 to 14 and different substituents can be
attached to the aromatic ring in the {\it para} position (position 2 in
Fig.1) and at the phenolic oxygen. During the last decades calix[n]arene
have been extensively utilised as molecular platforms for the synthesis of
potent receptors for neutral and charged species and for their ability to
form endo-cavity inclusion complexes with guests of complementary size.\cite
{Mandolini,Asfari} On the other hand the discovery of quantum tunnelling
processes involving the hydrogen atoms of the OH groups was recently
reported. \cite{Brougham,Horsewill} Both the design of molecules with an
internal cavity able to recognize metal cations and/or neutral molecules 
\cite{Mandolini,Asfari} and the study of the transfer of protons involved in
hydrogen bonding, which is fundamental to many chemical and biological
processes \cite{Liu,MacGilli}, are important subjects; however, many
questions remain unsolved and, especially concerning the second item, very
few studies have been reported for the condensed state.

Quite recently NMR relaxometry measurements indicated that hydrogen atoms in
calix[4]arene and p-tert-butyl-calix[4]arene might move both by means of a
thermally activated hopping process and by a coordinated quantum tunneling
process by which the four H atoms of the O-H groups move in a coordinated
fashion giving rise to an interconversion between the two tautomers of the
molecule. \cite{Brougham,Horsewill} In p-tert-butyl-calix[4]arene the
coordinated proton transfer (relaxation) rate , $\tau _{c}$, has been
directly determined and it was found to be independent of temperature
between 15 and 19 K, clearly demonstrating that the dynamical behavior of
the system is non-Arrhenius in character. \cite{Brougham} In the
calix[4]arene molecule, in addition, a thermally activated proton hopping
process was found, with a classical Arrhenius behavior described by a
relaxation time $\tau _{c}$=1.2$\cdot $10$^{-12}\cdot $e$^{\frac{900K}{kT}}$
s, which has been ascribed to rotational flips of the single O-H group (jump
of the H atom between state 1 and 1' in Fig.1). \cite{Horsewill}

In the following we will report anelastic spectroscopy measurements on
calix[4]arene containing four phenolic units (Fig.1). All the carbon and
oxygen bonds are saturated by hydrogen atoms. We will show that a peak in
the anelastic spectrum can be possibly ascribed to the same hopping process
detected by NMR measurements and we will discuss in detail a quantitative
analysis of this peak.

NMR relaxometry is a well-established experimental technique for the study
of molecular dynamics. From the present study we can deduce that anelastic
spectroscopy has great potentialities to study the movements of atoms in
molecules at a mesoscopic (several-molecule) level.

\section{EXPERIMENTAL AND RESULTS}

Calix[4]arene is produced as a powder sample. In order to measure its
elastic properties, the powder was mixed with a pure KBr powder and was
pressed at high pressure in order to obtain a bar of 7x38x2.0 mm$^{3}$. A
pure KBr bar was prepared in the same way and its anelastic spectrum was
also measured for comparison.

The complex Young's modulus $E\left( \omega \right) =E^{\prime }+iE^{\prime
\prime }$, whose reciprocal is the elastic compliance $s=E^{-1}$, was
measured as a function of temperature by electrostatically exciting the
lowest odd flexural mode. The vibration amplitude was detected by a
frequency modulation technique. The vibration frequency, $\omega /2\pi $, is
proportional to $\sqrt{E^{\prime }}$, while the elastic energy loss
coefficient (or reciprocal of the mechanical $Q$) is given by\cite{Nowick} $%
Q^{-1}\left( \omega ,T\right) =$ $E^{\prime \prime }/E^{\prime }=$ $%
s^{\prime \prime }/s^{\prime }$, and was measured by the decay of the free
oscillations or the width of the resonance peak. The imaginary part of the
dynamic susceptibility $s^{\prime \prime }$ is related to the spectral
density $J_{\varepsilon }\left( \omega ,T\right) =$ $\int dt\,e^{i\omega
t}\left\langle \varepsilon \left( t\right) \varepsilon \left( 0\right)
\right\rangle $ of the macroscopic strain $\varepsilon $ through the
fluctuation-dissipation theorem, $s^{\prime \prime }\propto $ $\left( \omega
/2k_{\text{B}}T\right) J_{\varepsilon }$.

In order to study the presence of thermally activated peaks in the anelastic
spectrum of calix[4]arene the measurements are carried out at different
vibration frequencies. Higher order flexural modes could not be exited on
the sample presently used; however, as the flexural mode frequency, $\omega
/2\pi $, is directly proportional to the height of the sample, $h$, the
specimen thickness was reduced with sand paper changing $h$ from 2.2 to 1.1
mm and consequently $\omega /2\pi $ changed by almost a factor 2.

The anelastic spectra of the pure KBr and the mixed KBr-calix[4]arene bar
are reported in Fig.2. In the temperature range from 1.1 to 400 K the KBr
sample presents just a featureless background which increases slightly
between 1 and 300 K and more significantly above room temperature. Instead,
the calix[4]arene sample exhibits three well developed peaks around 60 (peak
P1), 150 (P2) and 270 K (P3). As they are not present in the KBr specimen,
they can be attributed to intrinsic anelastic processes inside the
calixarene molecule.

An analysis of the three peaks is presented in the next section, where it
will be shown that the peak at the lowest temperature can be identified with
the same thermally activated hopping process of H atoms detected by NMR
measurements, giving rise to rotational flip of the OH bond. \cite{Horsewill}

\section{DISCUSSION}

In order to obtain useful information from the three peaks present in the
spectrum of calix[4]arene a quantitative analysis of them is needed. The
most common model describing relaxation processes (anelastic, dielectric,
etc.) is due to Debye. The contribution of a relaxation process to the
dynamic Young's modulus is: \cite{Nowick}

\noindent 
\begin{eqnarray}
\left[ 
\begin{array}{l}
Q^{-1} \\ 
\delta E^{\prime }/E^{\prime }
\end{array}
\right] &=&Ev_{0}c\left( \lambda _{1}-\lambda _{2}\right) ^{2}\cdot 
\nonumber \\
&&\cdot {\frac{f_{1}f_{2}}{T}}{\frac{1}{1+\left( \omega \tau \right)
^{2\alpha }}}\left[ 
\begin{array}{l}
\left( \omega \tau \right) ^{\alpha } \\ 
-1
\end{array}
\right]  \label{eq:1}
\end{eqnarray}

\noindent where $c$ is the molar concentration of relaxing units; $f_{1}$
and $f_{2}$ are the equilibrium fractions in the two possible configurations 
$1$ and $2$ between which the relaxation can occur with a characteristic
time $\tau \left( T\right) $ which classically follows an Arrhenius law, $%
\tau =\tau _{0}\exp (E_{a}/kT)$; $\lambda _{i}$ is the elastic dipole of the
state $i$, i.e. the average macroscopic strain which the sample would have
if each cell (molecule) contained one relaxing unit in the state $i$; $v_{0}$
is the cell (molecule) volume and $\alpha \equiv 1$. The contribution of a
relaxation process to the absorption is a peak whose maximum is centered at
the temperature at which $\omega \tau =1$. Since $\tau $ decreases as $T$
increases, the peak shifts to higher temperature if measured at higher $%
\omega $. Around the same temperature, the real part of the Young's modulus
presents a negative step whose amplitude, from theoretical models, is twice
the peak height.

The Debye formula describes a relaxation process between two possible states
which occurs with a single activation energy, $E_{a}$, and a single
characteristic time, $\tau $. Often real processes are broader than Debye
peaks; in those cases a correction factor, $\alpha $, is introduced in (1)
(Fuoss-Kirkwood model \cite{Nowick}). Physically this means that the
relaxation process has a distribution of activation energies and relaxation
times which become broader as $\alpha $ decreases.

A first quantitative analysis of the $Q^{-1}$ spectrum of calix[4]arene was
obtained after the subtraction of a smooth background, adding a Debye
contribution for peak P1, a Debye contribution for peak P2 and a
Fouss-Kirkwood peak for peak P3. The experimental data and the theoretical
curves at the two measured frequencies are reported in Fig.3. The parameters
of the three peaks are reported in Table I.

P1 and P2 can be described by the Debye process characterized by a single
relaxation time, with the assumption that the $\tau \left( T\right) $
follows an Arrhenius law, $\tau =\tau _{0}\exp (E_{a}/kT)$. For peak P3 the
fit is less satisfactory and this may be due to an imperfect substraction of
the background, which in that range of $T$ starts to increase very fastly.
However P3 does not seem to be larger than a pure Debye relaxation with the
given parameters. For a Debye process, the height of the peak decreases as
the frequency increases almost as 1/T, unless the atom or the defect relaxes
between the states of an asymmetric potential well. In the present case for
both higher temperature peaks the 1/T law is not valid and a multiplicative
factor was introduced in (1) to account for the different height of the
peaks at the two different frequencies. In the following we will not analyze
the dependence of the peak heights on frequency, as the mechanical process
used to obtain a second vibrational frequency could have affected the
distribution of calixarene powders.

All the peaks are certainly due to some atomic motion inside the
calix[4]arene molecule as they are not present in the pure KBr specimen.
From Table I one can notice that the $\tau _{0}$ values for the three peaks
are the same within one order of magnitude. Peak P1 can be ascribed to a
relaxation process involving the reorientation of a single bond or
equivalently to the motion of a single atom. The similarity among the three
characteristic times may suggest that also the higher temperature relaxation
involves the motion of single atoms or a small number of them. Peak P3 is
not present in the anelastic spectrum of p-tert-butyl-calix[4]arene, whilst
a peak similar to P2 is present also in the last compound.\cite{Paolone}
Considering the similarities of the structure of the calixarene and
p-tert-butyl-calixarene molecules, it seems that peak P2 may be due to the
reorientation of the H atoms at positions 2 and 2' in Fig.1.\cite{Paolone}
However more experimental studies are needed to ascribe peak P2 and P3 to
some physical processes. We want to point out that the activation energy and
the relaxation time obtained from the fit procedure are not affected by the
possible uncertainty on the peak height.

In the following we will deal with peak P1. It can be quite satisfactorily
described by a Fuoss-Kirkwood model with $\alpha =0.48$; it means that the
characteristic time and activation energy have a continuous distribution
function around the values $\tau _{0}=0.8\cdot 10^{-12~}s$ and $E_{a}=1120~K$%
. These values are quite similar to the parameters of the thermally
activated proton hopping process, with a classical Arrhenius behavior
measured by NMR relaxometry, which gave $\tau _{0}=1.2\cdot 10^{-12}~s$ and $%
E_{a}=900~K$.\cite{Horsewill} This close similarity indicates that the NMR
and anelastic relaxation processes may be originated by the same physical
process, that is the rotational flips of the O-H groups of the calix[4]arene
molecules.\cite{Horsewill}

In order to gain insight and to have more information about the distribution
functions of $\tau $ and $E_{a}$, we tried a different approach based on a
direct evaluation of the distribution of activation energies; at this step
we fixed $\tau _{0}$ to the single value obtained for the Fuoss-Kirkwood
peak, both for simplicity of calculation and considering that we are dealing
with a single relaxation process which corresponds to a well defined
reorientation. This approach is very useful in describing a similar peak
found in p-tert-butyl-calix[4]arene, which cannot be modelled by a Debye or
by a Fouss-Kirkwood relaxation or even by the sum of a big number of them.%
\cite{Paolone} The expression for the elastic energy loss coefficient used
for the fit procedure was:

\begin{eqnarray}
Q^{-1} &=&a\int\nolimits_{0}^{\infty }{\frac{f_{1}f_{2}}{T}\cdot }  \nonumber
\\
&&\cdot \frac{\omega \tau \left( E_{a}\right) }{1+\left[ \omega \tau \left(
E_{a}\right) \right] ^{2}}g\left( E_{a}\right) dE_{a}  \label{eq:2}
\end{eqnarray}

\noindent where $a$ is a constant, $g\left( E_{a}\right) $ is a normalized
distribution function of the activation energy, $E_{a}$. Both a Lorentzian
and a Gaussian distribution function have been tested. The Gaussian
distribution is able to better reproduce the experimental data than a
Lorentzian curve. The best fit curves using a Gaussian distribution function
are reported in Fig.4. The best fit parameters are: $\tau _{0}=0.8\cdot
10^{-12~}s$, $E_{a}=1070~K$, $\sigma \left( E_{a}\right) =130~K$, where $%
\sigma \left( E_{a}\right) $ is the variance of the Gaussian distribution.

The existence of a distribution of activation energies more than a single $%
E_{a}$ can be explained considering that each H atom of the OH groups in a
calixarene molecule moves in an environment (potential well) which can vary
as the others three H atoms of the OH groups in the same molecule change
their position. Moreover the potential energy of the H atom can be at least
partially influenced by the surrounding calix[4]arene molecules.

A distribution of activation energies for a relaxation process measured by
NMR relaxometry, has been discussed also in the case of 1,3-di-{\it t}%
-butylbenzene, which was a very interesting case for the study of H dynamics
in organic molecules in the condensed state. \cite{Beckmann} The
experimental data were described by a non-Bloembergen-Purcell-Pound (BPP)
spectral density. In the case of anelastic relaxation, the BPP spectral
density has its exact counterpart in the Debye model with an Arrhenius-like
dependence of $\tau $ on $T$ and $E_{a}$. Beckmann {\it et al.} pointed out
that the non-BPP behavior can also be interpreted in terms of a distribution
of exponential correlation function (leading to the Arrhenius law), with the
distribution at least partially characterized by a correlation time $\tau $
which could either be a cutoff correlation time or a mean correlation time,
depending on the model. Correspondingly there is a distribution of
activation energies. However the relation between the two distribution
functions cannot be determined until the dependance of $\tau $ on $E_{a}$ is
modeled. In the present case $\tau _{0}$, which has been fixed to the best
fit value, can be interpreted as a mean correlation (relaxation) time.
Moreover the Arrhenius dependence of $\tau $ on $T$ and $E_{a}$ has been
explicitly used. However the Gaussian distribution function of the
activation energies has not been derived from a model, but has been found to
describe quite well the experimental data.

A last consideration regards the lack of any quantum tunneling process in
the anelastic spectrum of calix[4]arene, as that detected by NMR relaxometry.%
\cite{Horsewill} One possible explanation for this, is that we perform
measurements of the dynamical Young's modulus as a function of temperature,
measured with a practically fixed frequency. If the correlation time $\tau $
in the tunnelling process becomes temperature independent, like in the case
of p-tert-butyl-calix[4]arene, the condition $\omega \tau =1$ in order to
have a peak in the elastic energy loss coefficient could be never fulfilled.

\section{CONCLUSION}

We measured the anelastic spectrum of a bar of KBr containing a few percent
of calix[4]arene powder. We observed three thermally activated peaks. The
lowest temperature peak can be described both by a Fuoss-Kirkwood model and
by a model considering a Gaussian distribution of activation energies,
leaving the correlation time $\tau _{0}$ fixed. Due to the similarity
between the physical parameters of this peak and of a relaxation found in
the NMR spectrum of the same compound, the anelastic peak can be ascribed to
hopping of H atoms of the OH groups, which corresponds to a flip-flop of the
OH bond. From the results of the present study, anelastic spectroscopy
appears as a good tool for studying atomic motion inside molecules at a
mesoscopic (few molecules) level.


\begin{figure}[tbp]
\caption{Schematic representation of the calix[4]arene molecule. Black, grey
and white circles represent C, O and H atoms, respectively.}
\label{fig1}
\end{figure}

\begin{figure}[tbp]
\caption{The $Q^{-1}$ spectrum and the relative variation of the real part
of the Young's modulus as a function of temperature for the pure KBr and the
mixed KBr-Calix[4]arene bars.}
\label{fig2}
\end{figure}

\begin{figure}[tbp]
\caption{The $Q^{-1}$ spectrum of calix[4]arene measured at two different
frequencies (symbols) and the best fit curves (continous lines) obtained
adding two Debye contributions for the higher temperature processes and a
Fuoss-Kirkwood relaxation with $\alpha = 0.48$ for the lower temperature
process. In the inset the region around 60 K is reported in more detail.}
\label{fig3}
\end{figure}

\begin{figure}[tbp]
\caption{The $Q^{-1}$ spectrum of calix[4]arene measured at two different
frequencies (symbols) and the best fit curves (continous lines) obtained
adding two Debye contributions for the higher temperature processes and a
relaxation with the distribution of activation energies as discussed in the
text for the lower temperature process. In the inset the region around 60 K
is reported in more detail.}
\label{fig4}
\end{figure}

\begin{table}[tbp]
\caption{The best fit parameters of the three anelastic peaks of
calix[4]arene.}
\begin{tabular}{lcccc}
$peak$ & $T_{max} (K)$ & $\tau _{0}(s)$ & $E_{a}(K)$ & $\alpha$ \\ 
$P1$ & $60$ & $0.8\cdot 10^{-12}$ & $1120$ & $0.48$ \\ 
$P2$ & $150$ & $2.0\cdot 10^{-11}$ & $2370$ & $1$ \\ 
$P3$ & $270$ & $3.0\cdot 10^{-12}$ & $4720$ & $1$%
\end{tabular}
\end{table}


\begin{references}
\bibitem{Gutsche}  C. D. Gutsche, in ''{\it Monograph in Supramolecular
Chemistry}'', ed. J.F. Stoddart, The Royal Society of Chemistry, Cambridge,
1998.

\bibitem{Mandolini}  {\it Calixarenes in Action}, L. Mandolini, R. Ungaro
eds.; Imperial College Press, London, 2000.

\bibitem{Asfari}  {\it Calixarenes 2001}, Z. Asfari, V. Bohmer, J.
Harrowfield, J. Vicens eds.; Kluwer Academic Publishers, Dordrecht, 2001.

\bibitem{Brougham}  D.F. Brougham, R. Caciuffo and A.J. Horsewill, Nature 
{\bf 397}, 241 (1999).

\bibitem{Horsewill}  A.J. Horsewill, N.H. Jones and R. Caciuffo, Science 
{\bf 291}, 100 (2001).

\bibitem{Liu}  K. Liu {\it et al.}, Nature {\bf 381}, 501 (1996).

\bibitem{MacGilli}  L.R. MacGillivray and J.L. Atwood, Nature {\bf 389}, 469
(1997).

\bibitem{Nowick}  A.S. Nowick and B.S. Berry, {\it Anelastic Relaxation in
Crystalline Solids}, Academic Press, New York, 1972.

\bibitem{Paolone}  A. Paolone, R. Cantelli, R.G. Caciuffo, F. Ugozzoli,
unpublished.

\bibitem{Beckmann}  P.A. Beckmann, A.I. Hill, E.B. Kohler and H. Hu, Phys.
Rev. {\bf B 38}, 11098 (1988).
\end{references}
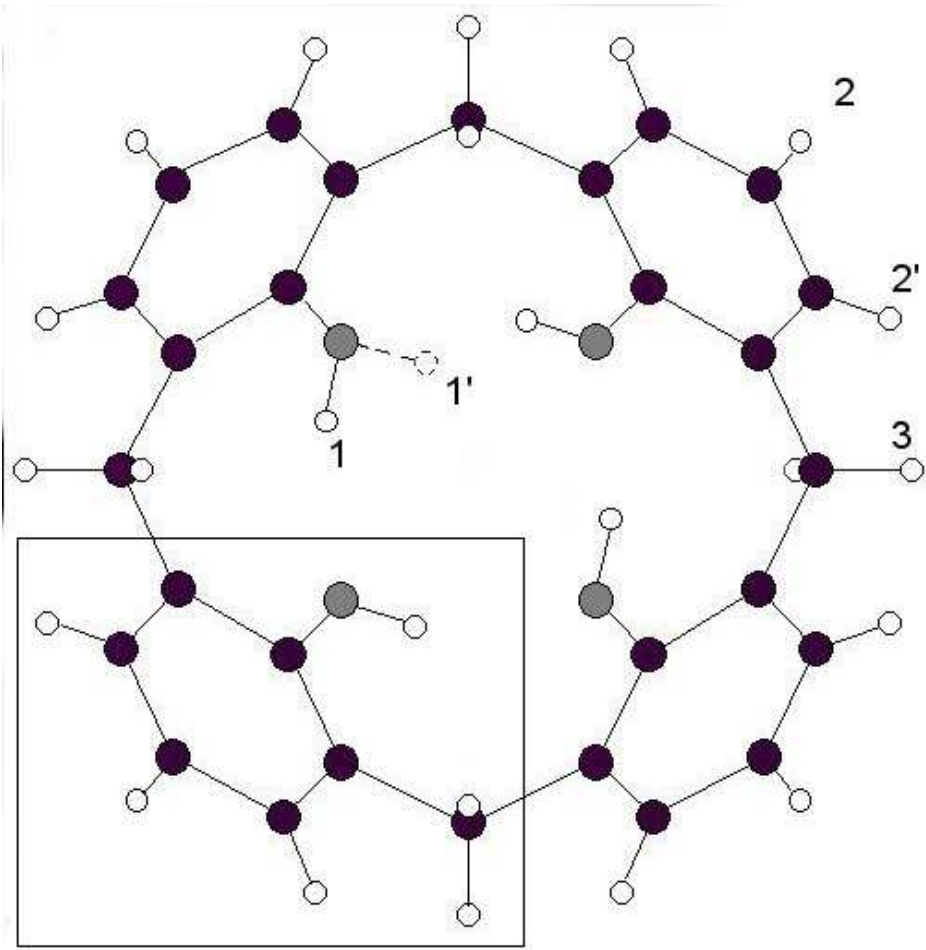
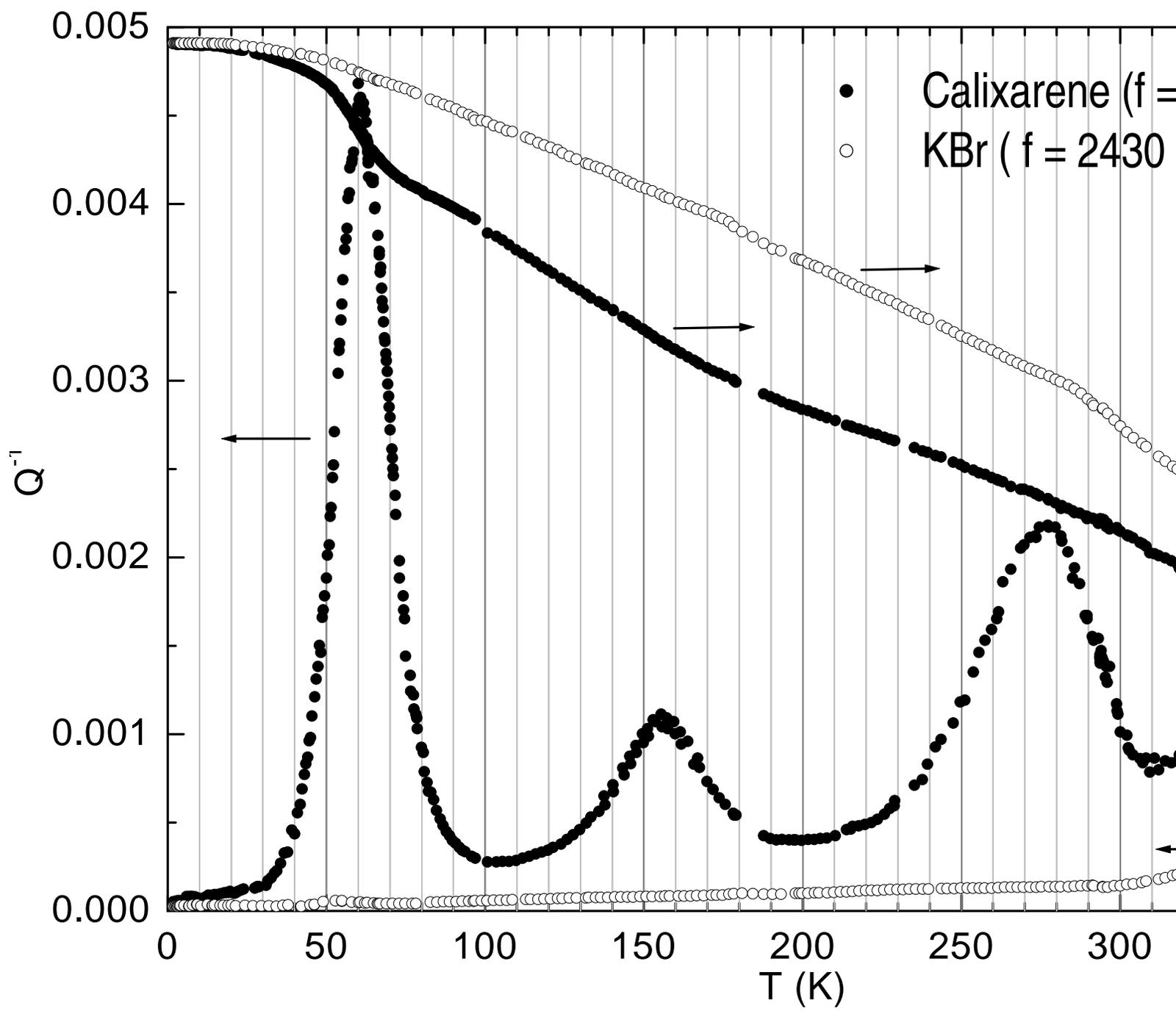
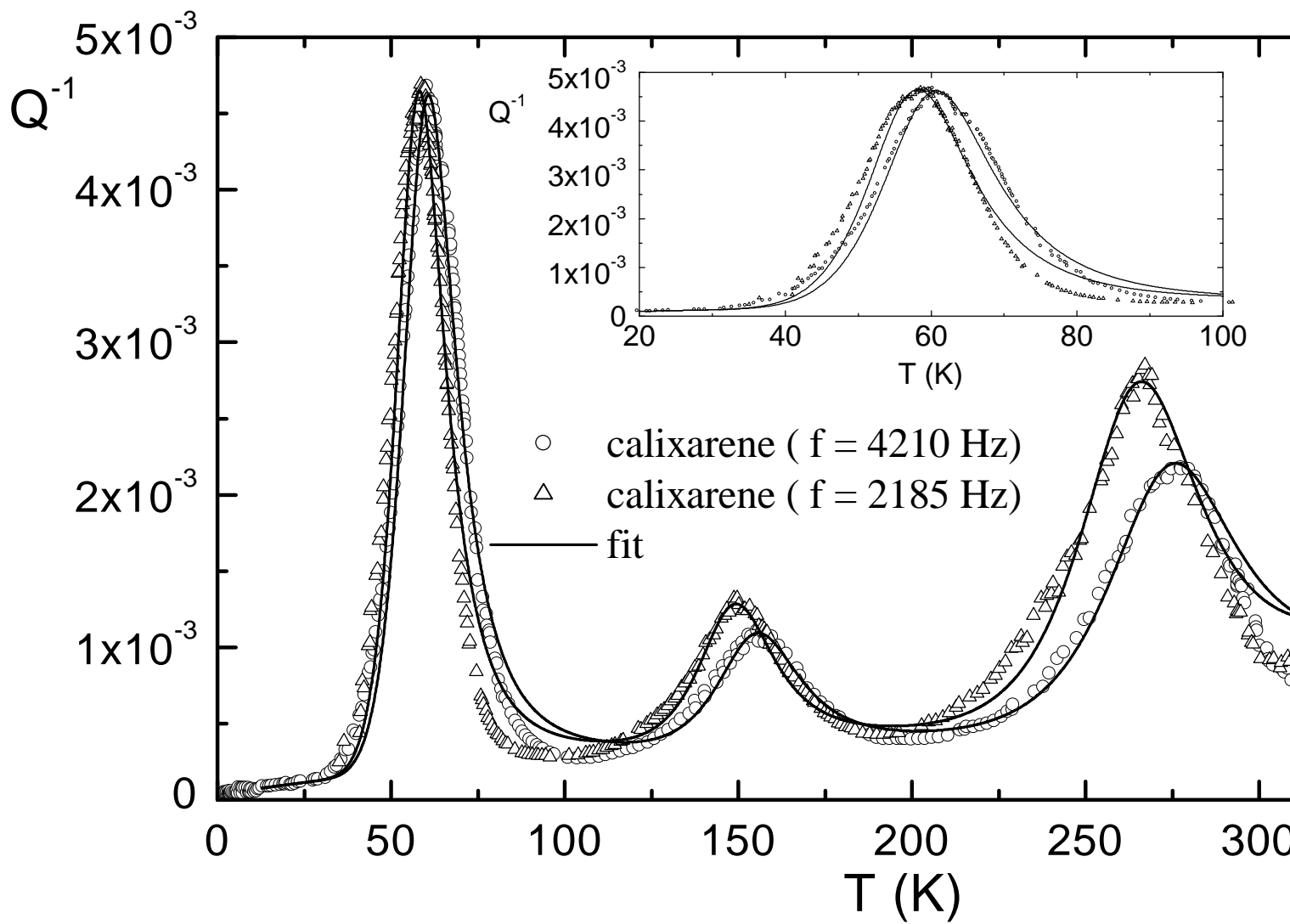
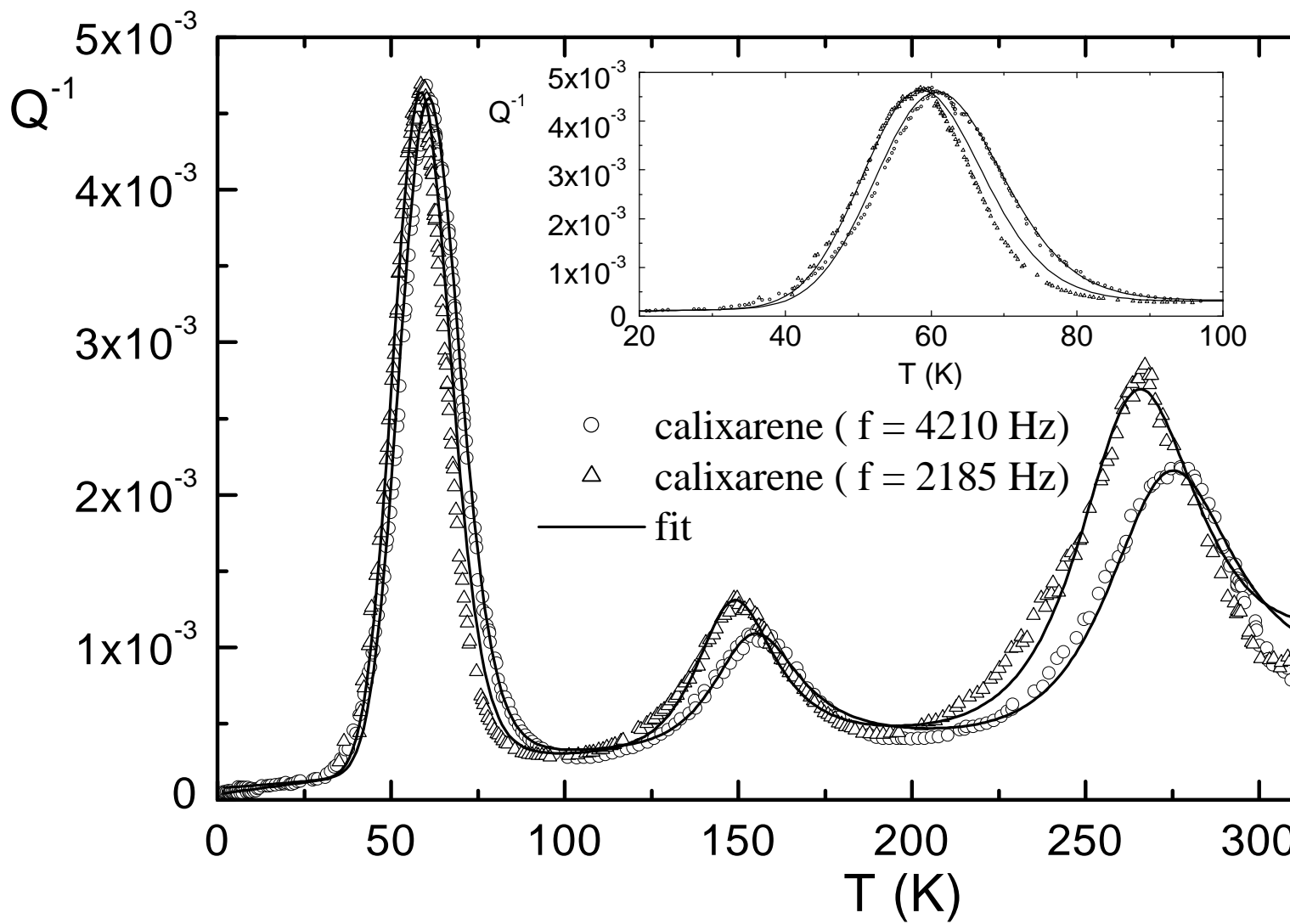
\end{document}